\documentclass[12pt]{iopart}

\usepackage{iopams}
\renewcommand{\vec}{\boldsymbol}
\usepackage{epsf}
\usepackage{graphicx}
\begin{document}

\title[]{Laser in axial electric field as a tool to
search for P-, T- invariance violation}

\author{V G Baryshevsky, S L Cherkas and D N Matsukevich\footnote{Present address: School of Physics,
Georgia Institute of Technology, Atlanta, Georgia 30332-0430.}}

\address{Institute for Nuclear Problems, 11 Bobruiskaya str.,
220050 Minsk, Belarus}
\begin{abstract}
We consider rotation of polarization plane of the laser light when
a gas laser is placed in a longitudinal electric field (10~kV/cm).
It is shown that residual anisotropy of the laser cavity $10^{-6}$
and the sensitivity to the angle of polarization plane rotation
about $10^{-11}-10^{-12} $ $rad$ allows one to measure an electron
EDM with the sensitivity about $10^{-30}$ $e \times cm$.
\end{abstract}

\maketitle

\section {Introduction}

 The standard model predicts the dipole moment of the electron at a
 level of about
 $10^{-40}$ $e \times cm$ while some variants of supersymmetric models
 forecast $10^{-30}$ $e \times cm$ \cite{ellis,masina,peskin}. Since the supersymmetry is an
 important ingredient of modern physics it would be desirable to
 achieve the sensitivity of measurements enabling us to test these
 predictions. The present limit on the electron EDM is $d_e<1.6\times
 10^{-27}$ \cite{reg}.

 The measurement of an angle of the
polarization plane rotation of light when it propagates through a
gas immersed in an electric field is one of the possible ways of
 searching  the electric dipole moment of an electron. The interaction of
 an
 electric field with the electron dipole moment leads to the splitting
 of the atomic levels, analogous to the Zeeman effect and,
 consequently,
 to the polarization plane rotation (similar to Faraday effect) when
 a photon propagates along the electric field direction.

\begin{figure}[h]
\vspace{0. cm} \hspace{2.5 cm}
\includegraphics[width=3.9cm]{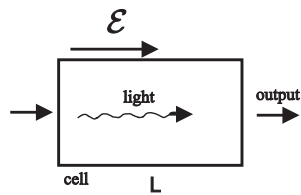} \caption{Scheme of the
transmission experiment.} \label{cell}
\end{figure}

It was shown in \cite {bar,mats,bar1} that in addition to the
atomic level splitting one more mechanism leading to the light
polarization plane rotation exists. It is the interference of the
Stark and P-T- invariance violating transition amplitudes.

\begin{figure}[h]
\vspace{0.7 cm} \hspace{2.5 cm}
 \includegraphics[width=8. cm]{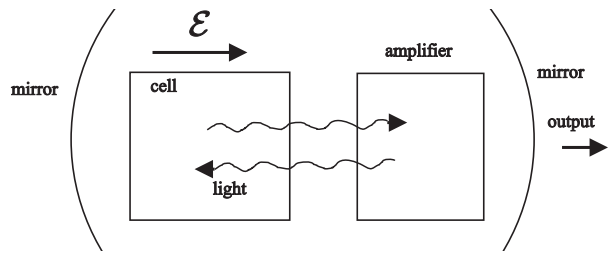}
 \vspace{. cm}
\caption{ A gas cell with an amplifier \cite {bar1} for
observation of the P- T- odd polarization plane rotation.}
\label{amplif}
\end{figure}

In a typical transmittance experiment (Fig. {\ref{cell}}) with a
gas cell the intensity of a light beam decreases when it
propagates in a medium. This restricts  the length available for
polarization plane rotation measurements \cite{hrip}. An idea to
use a photon trap (resonator) with an amplifier (Fig.
\ref{amplif}) to compensate light absorption was proposed in \cite
{bar1}. Because the trap contains exited medium, the amplification
cancels the losses  and light can stay in the resonator for a long
time.

\begin{figure}
\vspace{0.1cm} \hspace{3.5 cm}
  \includegraphics[width=4.5cm]{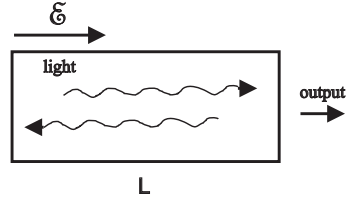}
\vspace{0. cm} \caption{Laser placed in the external electric
field.} \label{pin}
\end{figure}

The simplest type of the trap is a laser placed in the electric
field (Fig. \ref{pin}). This system is similar to the laser in the
magnetic field considered in 70th \cite{lamb,gal,harr,gal1,voit}
and recently as a footing of the laser magnetometers \cite
{bret,bret1}.

 The main difficulty for laser magnetometery is linear
anisotropy of losses of the resonator \cite {bret,bret1}. The same
difficulty is deferred on the measurements of P-T- noninvariant
rotation in the electric field, because the linear anisotropy of
losses  of the resonator
 is much greater than the circular anisotropy created by the
electric field provided that P- and T- invariance breaks. It
should be remembered that in the absence of the linear anisotropy
of losses an angle between the polarization plane  of the laser
radiation and some axis increases linearly with time, and the
angular velocity of the polarization plane rotation is
proportional to the P- T- odd polarizability of the atoms and
strength of the electric field.
    Large linear anisotropy
forbids  polarization plane rotation. When the electric field
turns on, the polarization plane rotates for a small angle and
then stops. Only this angle should be measured. Below we give a
detailed theoretical  analysis of such a photon trap
 for measurements of the P-T- noninvariance.

\section {The angle of polarization plane rotation in the
electric field}

Let us consider a stationary electromagnetic wave in a resonator
containing an exited medium possessing linear and circular
anisotropy.
 The effect of polarization plane rotation can be described by
 the P- T- noninvariant term $n ^
 {ij}_{PT} \sim e ^ {ijk} \mathcal E^k $ in the tensor of the
 refractive index of the medium \cite {zel,flam}, where $e^{ijk}$ is
 the
 completely antisymmetric tensor and $\vec {\mathcal E}$ is the
 external electric field.
Let the  mirrors of the resonator be perpendicular to the
$z$-axis, and the external electric field $\vec {\mathcal E}$ be
directed along it.

It is convenient to choose the reference frame providing for the
matrix of anisotropic
 losses of the resonator to be diagonal.
Thus the tensor part of the generalized refractive index (see
Appendix), which includes the resonator, has the form:
\begin{equation}
 \Delta \hat n^\prime\ = \left (
\begin {array}{cc}
-d-i\chi\;&b+ia \\
b-ia\;&d+i\chi
\end {array}
\right),
\end{equation}
where $\chi$, $a$ describe linear anisotropy of losses and
circular phase anisotropy correspondingly.  Linear phase
anisotropies are given by $b$ and $d$. Prime marks the refractive
index of the exited medium  to distinguish it from that for the
medium in the ground state. The refractive index acts as matrix in
the space of the vectors
 $\left(
\begin {array}{c}
E_x \\
E_y
\end {array}
\right) $, where $E_x$ and $E_y$ are the components of  the
electric field of the electromagnetic wave.

Linear anisotropy of losses $ \chi =\frac{1}{4}\left(\frac {1}
{Q_y}-\frac {1} {Q_x}\right)\approx \frac {1} { 4 } \frac {\Delta
Q} {Q^2} $, where $ \Delta Q=Q_x-Q_y $ (see Appendix). Here $Q_x$
and  $Q_y$ are the finesse of the resonator for the light
polarized along the $x$ and $y$ axes, respectively. In principle,
very low anisotropy of losses can be achieved. For instance, in
the experiments \cite {bret,bret1} the quantity  $ \frac {\Delta
Q} {Q} =\frac{Q_x-Q_y}{Q}\sim 10 ^ {-5}$. In our estimates we  use
the value of $10^{-6}$ in a hope that the progress in the
technology of unstressed materials and the resonator design will
allow it to be reached.

Circular phase anisotropy $a =\frac {1 }{ 2 } \Delta n
_{PT}^\prime = \frac {1 }{ 2 } (n ^\prime_+-n_-^\prime) $ is
produced  by the electric field if P- T- invariance is violated.
Linear phase anisotropies $d=\frac {1 }{ 2 } (n
^\prime_y-n_x^\prime)$ and $b=\frac {1 }{ 2 } (n
^\prime_{135^o}-n_{45^o}^\prime)$, where $n ^\prime_x, n
^\prime_y$, $n ^\prime_{135^o}$, $n_{45^o}^\prime$ are the
refractive indexes for the wave polarized along the $x$-axis and
$y$-axis, at the angle $135^o$ and at the angle $45^o$ (relative
to the $x$-axis), respectively. If one rotates the reference frame
at ${45^o}$, the quantity $b$ takes up positions on the diagonal
of the refractive index matrix. However more convenient is the
circular basis $
 \left (
\begin {array} {c}
E_+\\
E_-
\end {array} \right)
=\frac {1} {\sqrt {2}}\left (
\begin {array} {c}
- E_x-iE_y\\
E_x-iE_y
\end {array} \right) $ in which the refractive index takes the form:
\begin{equation}
 \hat\Delta n^\prime\ = \left (
\begin {array}{cc}
a\;&d-ib+i\chi \\
d+ib+i\chi\;&-a
\end {array}
\right).
\end{equation}

Measurements should be performed by the analyzer placed
perpendicularly to the polarization plane of the laser beam (when
the electric field is turned off). The orientation of the
polarization plane of the laser light is determined by residual
anisotropy of losses in the resonator  and coincides with one of
the main axes i.e. it is directed along the vector $ \vec h =
\{\chi, 0,0 \}$.

As the electric field is  turned on, the circular anisotropy $a$
depending linearly on the electric field  appears and the
polarization plane rotates by the angle $ \phi_{las} $, which can
be found from the Eq. (\ref{formap}) given in appendix. This
equation
 describes stationary polarization
of light. The angle $ \phi_{las} $ is the half of the angle
between vector $\vec O$ at $a=0$, when there is no electric field
and that at some $a$ produced by the turned on electric field.
  In a typical
experimental situation $a, b, d\ll\chi$. Vectors $\vec h $ and
$\vec r$ in (\ref{formap}) are $ \vec h= \{\chi, 0,0 \}  $ (we
omit circular dichroism due to an electric field for simplicity)
and $ \vec r = \{d, b, a \} $. The stationary solution for the
vector $\vec O$ describing polarization state of the standing
electromagnetic wave according to (\ref{formap}) has the form:
\begin {equation}
\vec O_0 =\left (1,
    \frac {a} {\chi} + \frac {bd} {\chi^2}, \frac {ad} {\chi^2}-
      \frac {b} {\chi} \right)
\label {xi0}
\end {equation}
in the first order on $a, b, d $.

 Eq. (\ref {xi0}) gives the angle of polarization plane rotation
\begin {equation}
\phi _ {las} \approx\frac{1}{2} \frac {O _ {0y}} {O _ {0x}}
\approx \frac {a} {2\chi} \approx {\Delta n _ {PT} ^ \prime} \frac
{Q^2} {\Delta Q}. \label {sng}
\end {equation}

As we can see from (\ref{xi0}) the additional parasitic angle $
\frac{bd}{2\chi^2} $ appears. However, this angle can not depend
linearly on the electric field. One more additional angle
(so-called "base angle" \cite{hrip}) arises due to inexact
perpendicular orientation of the analyzer with respect to the
polarization plane of the laser beam.

The conventional experimental method implies  modulation of the
"base angle" with the frequency $ \Omega $ by the additional
Faraday element placed between laser and analyzer. The signal at
the output of analyzer is proportional to the squared sum of the
P-T- violating angle of rotation and the "base angle". The
presence of $ \Omega $ component in the Fourier transform of the
output signal is the signature of P-, T- invariance violation.

\section {Estimates for the "trap" and "transmittance" layouts}

Let us estimate the advantage of the laser experiment compared to
the light transmission experiment using a cell (Fig. \ref{pin}).

The P-, T- violating refractive index does not depend on the type
of the atomic transition, i.e. it is approximately the same for
the dipole electric, magnetic and strongly forbidden magnetic
transitions \cite{mats}. In the  transmittance experiments with a
cell the angle of rotation is usually measured for two absorption
lengths, thus, it is reasonable to choose transitions with the
greatest absorption length. These are magnetic  dipole and
strongly forbidden magnetic transitions. The angle of polarization
plane rotation in a cell for the length $L$ is equal to (see
\cite{bar,mats,bar1,hrip}):
\begin {equation}
 \phi=\frac{1}{2}\Delta
n_{PT}k L,\label{phil}
\end {equation}
where $k$ is the wave number. As we have mentioned, $\Delta
n_{PT}$ in the equation (\ref {phil}) differs from  $\Delta
n_{PT}^\prime$ in the expression (\ref{sng}). The first quantity
corresponds to a  medium in the ground state and transitions
happen from the ground level to the exited one whereas, in the
case of the laser medium, the transitions happen from a top level
to the bottom.  These quantities are connected to each other by
the relation
\begin {equation}
\Delta n _ {PT} ^ \prime =\frac {\Delta N} {N} \Delta n _ {PT},
\label {inv}
\end {equation}
where $N $ is the concentration of atoms, and $ \Delta N $ is the
density of inversely populated atoms. Substituting $ \Delta n _
{PT} ^ \prime $ from (\ref {inv}) to the equation (\ref {sng}) one
obtaines:
\begin {equation} \phi_{las}\sim  \frac {\Delta N} {N}
\Delta n _ {PT} \frac {Q^2} {\Delta Q} \sim  \Delta n _ {PT}
k\frac {1} {N\sigma} \frac {Q} {\Delta Q}. \label {result}
\end {equation}
In the derivation of the latter equation we have taken into
account the condition of laser operation
\begin{equation}
\Delta N \sigma =\frac {k} {Q}, \label{lasf}
\end{equation}
where  $ \sigma $ is the absorption cross section for this
transition. From the equations (\ref{result}) and (\ref{phil}) we
can see that the angle of polarization rotation in the laser is
equal to the angle of rotation at the absorption length $L_{abs}=
\frac {1} {N\sigma} $ multiplied by $ \frac {2 Q} {\Delta Q} $.
Thus one expect to obtain $ \frac {2 Q} {\Delta Q} \sim 2\times 10
^ {6} $ enhancement in comparison with a layout, using a cell. Let
us remind that the experiment with a cell uses the strongly
forbidden  magnetic transition (i.e. transition between shells
with different main quantum numbers), therefore the real gain will
only appear if the laser also operates at transitions of this
type.  In the laser operating at ordinary electric dipole
transitions, a very low inversion of population is required to
compensate absorptions in the resonator, because $\sigma$ in
(\ref{lasf}) is large. The real part of the refraction index is
also proportional to the inversion of population and additional
suppression given by (\ref{inv}) arises. Thus, the most of the gas
lasers, using $E1$ transitions are unsuitable as a trap for
measuring P-T- violation. Although there are no lasers working at
a strongly forbidden magnetic transition, lasers working at a
magnetic dipole transition do exists. One of such example, namely,
chemical iodine laser will be considered below.

The P- T- odd refractive index can be expressed in terms of the P-
T-odd polarizability $ \beta _ {PT} $ of an atom:
\begin{equation}
\Delta n _ {PT} =-4\pi N \beta _ {PT}.
\label{np}
\end{equation}
Two
mechanisms contributing to $\beta_{PT}$ were considered in
\cite{bar,mats,bar1}. The first one is the interference of Stark
and P-T- odd transition amplitudes. The value of $\beta_{PT}$ in
this case can be estimated as:
\begin{equation}
\beta^{mix}_{PT}\sim\sum_{m,n}\frac{<g|H_T|m><m|d^j|c><c|d^j|n><n|\vec
d\vec {\mathcal E}|g>
}{(\varepsilon_m-\varepsilon_c)(\varepsilon_g-\varepsilon_c
+\omega+i{\Gamma}/{2})(\varepsilon_n-\varepsilon_g)}, \label{bmix}
\end{equation}
where $\omega$ is the laser working frequency, corresponding to
the resonator own frequency; $m, n$ are some intermediate atomic
levels, $\varepsilon_g, \varepsilon_c, \varepsilon_n,
\varepsilon_m$ are the energies of the levels, $d^j$ are the
components of the operator $\vec d$ of the atom dipole moment
(summation on $j$ is implied in (\ref{bmix}) and further), $H_T$
is the operator of P- T-violating interaction. We assume that
$c\rightarrow g$, is the laser working
 transition, and $g$ is the ground state.

The dependence of polarizability on frequency is given by a
multiplier $\frac {1} {\omega-\omega_0+i\Gamma/2}$, where $
\omega_0 ={\varepsilon_c-\varepsilon_g} $ is the frequency of
transition and $ \Gamma $ should be read as denoting the recoil
line width.  To take into account Doppler broadening in a gas we
should average the multiplier over the Maxwell distribution of
atom velocities. According to ref. \cite{hrip} this reduces to:
\begin {equation}
\langle\frac {1} {\omega-\omega_0+i\Gamma/2}
\rangle\Rightarrow\frac {1} {\Delta_D} \left (g (u, v)-if (u, v)
\right), \label{avr}
\end {equation}
where $ \Delta_D=\frac{\omega}{c}\sqrt{\frac{2k_b\,T}m} $ there is
the Doppler line width, $c$ is speed of light, m is an atom mass,
$k_b$ is the Boltzman constant, $T$ is the temperature, $v =\frac
{\Gamma} {2\Delta_D} $, $u =\frac {\omega-\omega_0} {\Delta_D} $
is the detuning and
\[
\begin {array}{c}
g(u,v) \\
f(u,v)
\end {array}
\Biggl\}=
\begin {array}{c}
\mbox{Im} \\
\mbox{Re}
\end {array}\Biggl\}\sqrt{\pi}e^{-w^2}(1-\Phi(-iw)),
\]
$w=u+iv$, $\Phi(z)=\frac{2}{\sqrt{\pi}}\int_0^z dt e^{-t^2}$.

Using the above averaging in  (\ref{bmix})  we come to the
estimate:
\begin{equation}
\beta^{mix}_{PT}\sim\frac{<d>^3{\mathcal E}_z<H_T>}{(\Delta
\varepsilon)^2}\frac{g(u,v)}{\Delta_D}, \label{12}
\end{equation}
where $<d>$ is the typical value of the matrix element from the
operator of atom dipole moment , $\Delta \varepsilon\sim Ry$ ($Ry$
is Rydberg constant) is the typical value of difference in atomic
levels energies, ${\mathcal E}_z$ is the longitudinal component of
$\vec {\mathcal E}$. The above estimate of $\beta_{mix}^{PT}$ is
valid for all kinds of atomic transitions \cite{mats}: electric
dipole, magnetic dipole and strongly forbidden magnetic dipole.

The second mechanism suggests that the P-T- odd polarizability is
produced by the atomic levels splitting in the electric field due
to the atomic EDM. This leads to the estimates \cite{mats}:
\begin{eqnarray}
\fl\beta^{edm}_{PT}\sim \sum_{m,n} \frac{<g|\vec d\vec {\mathcal
E}|m><m|d^j|c><c|d^j|n><n|\vec d\vec {\mathcal
E}|g>}{(\varepsilon_m-\varepsilon_c)(\varepsilon_n-\varepsilon_g)}
 \nonumber\\~~~~~~~~~~~~~~~\times\vec
d_{at}\vec {\mathcal E} \,\frac{\partial}{\partial
\omega}\,\frac{1}{\varepsilon_g-\varepsilon_c+\omega+i\Gamma/{2}}
\label{bedm}
\end{eqnarray}
for the strongly forbidden magnetic transition and
\begin{equation}
\beta^{edm}_{PT}\sim  <g|\mu^j|c><c|\mu^j|g>\vec d_{at}\vec
{\mathcal E}\,\frac{\partial}{\partial
\omega}\,\frac{1}{\varepsilon_g-\varepsilon_c+\omega+i\Gamma/{2}}
\label{beds}
\end{equation}
for the magnetic dipole transition. Here $\mu^j$ are the
components of the  operator of atom magnetic moment, $d_{at}$ is
the dipole moment of the atom, which can be estimated as
$d_{at}\sim \frac{<d><H_T>}{\Delta \varepsilon}$. Averaging
(\ref{beds}) we obtain
\begin{equation}
\beta^{edm}_{PT}\sim \frac{<d>^4 {\mathcal E}_z^3d_{at}}{(\Delta
\varepsilon)^2}\frac{1}{\Delta_D^2}\frac{\partial g(u,
v)}{\partial u}  \sim \frac{<d>^5{\mathcal E}_z^3}{(\Delta
\varepsilon)^3}\frac{<H_T>}{\Delta_D^2}\frac{\partial g(u,
v)}{\partial u}
\end{equation}
for the strongly forbidden transition and
\begin{equation}
\beta^{edm}_{PT}\sim <\mu>^2 d_{at} {\mathcal
E}_z\frac{1}{\Delta_D^2} \frac{\partial g(u, v)}{\partial u}
 \sim \alpha^2\frac{<d>^3{\mathcal
E}_z<H_T>}{\Delta \varepsilon\,\Delta_D^2}\frac{\partial g(u,
v)}{\partial u}
\end{equation}
for the magnetic dipole transition, where $\alpha=\frac{e^2}{\hbar
c }$ \cite{lan4} is the fine structure constant, $<\mu>\sim \alpha
<d>$, $<d>\sim e\, a_0$, $a_0$ is Bohr radius.

Sources of P- T- violation are the electron EDM, EDM of the
nucleons, and the P-T-odd electron-nucleon interaction
\cite{mats}. For definiteness we consider only the first one. The
matrix element of P-T- odd interaction between atomic states can
be estimated as $<H_T>\sim 150 \frac{d_e}{<d>}\Delta \varepsilon$
\cite{mats}. This implies that the atom EDM is of the order of
$d_{at}\sim<d>\frac{<H_T>}{\Delta \varepsilon}\sim 150 d_e$. The
above estimation takes into account the Shiff theorem stating that
in nonrelativistic quantum mechanic atom EDM should be zero and
only relativistic effects  allows it to appear \cite{hrip}.
Relativistic effects are given by the multiplier ${\mathcal V
}^2/c^2\sim Z^2\alpha^2 $ \cite{hrip}, where $\mathcal V$ is the
typical electron velocity in atom, $Z$ is the atomic number.
However, the Shiff theorem does not concern the "mixing"
mechanism, because an atom EDM does not appear in it. Thus we have
to use two different $<H_T>$ to describe "mixing"  and "splitting"
mechanisms. Unfortunately, in Ref. \cite{mats} we used the single
$<H_T>$ with relativistic suppression  for both the mechanisms. In
this paper will take $<H_T>\sim 150 \frac{d_e}{<d>}\Delta
\varepsilon$ for the "splitting" mechanism and $<H_T>\sim
\frac{150}{Z^2 \alpha^2} \frac{d_e}{<d>}\Delta \varepsilon$ for
the "mixing" mechanism.

 The
 absorption cross section
  is given by
\begin {equation}
\sigma=\pi A \frac {c^2} {\omega^2} \frac {f (u, v)} {\Delta_D},
\label{cross}
\end {equation}
where $A$ is the Einstein coefficient of a given transition.  The
angle of polarization plane rotation at one absorption length $
\frac {1} {N\sigma} $ equals
\begin {equation}
\phi (L_{abs}) = \frac {2\pi\,\omega \beta_{PT}} {c\, \sigma}.
\end {equation}

We are coming now to the concrete systems.

\subsection{Iodin laser, operating on M1 transition}

 For lack of laser on strongly forbidden magnetic transition we
consider chemical atomic iodine gas laser employing
$^2P_{3/2}\rightarrow {^2P}_{1/2}$ ($\lambda=1.315 ~\mu m$)
magnetic transition \cite{gas,bred}, for witch $A=7.7 ~ c ^ {-1}$.
 According to
 eq. (\ref{12} ) P-T-odd polarizability for "mixing"
mechanism can be expressed as
$\beta_{PT}^{mix}=B_{PT}^{mix}\frac{g[u,v]}{\Delta_D}$, where
$B_{PT}^{mix}$ is determined only by the atom properties and
strength of external electric field but not the detuning and line
broadening. For iodin atom the above estimates give
$B_{PT}^{mix}=4.5\times 10^{-33}~cm^3 s^{-1}$ at ${\mathcal
E}=10^4~V/cm$.
 The angle of polarization plane rotation at one
absorbtion length and that for the laser system
$\phi_{las}=\phi(L_{abs})\frac{2\,Q}{\Delta Q}$
 are given in
Table \ref{tab1}.

Two first lines are associated with the top-table chemical iodine
lasers \cite{i1,i2,i3,i4} using chemical excitation of the Iodin
atoms by singlet oxygen:
\[
I+O_2(^1\Delta)\leftrightarrows I^*+O_2(^3\Sigma),
\]
where singlet oxygen is generated outside the laser and is
injected into the laser cavity together with the iodine. For our
case a design of reagent injecting and output have to be as
possible as axially symmetric to avoid transverse anisotropy of an
active medium.

Typically lasers of that type works at temperature $60-80~C^o$
(this promise low thermal drifts during measurements), pressure of
singlet oxygen $p\approx 1 ~torr$, and pressure of the iodine
$p_{[I^*]}\approx 10^{-2}p$. Almost all Iodine atoms are in the
exited state, because equilibrium in the above chemical reaction
is strongly accented to the right due to excess of
$O_2(^1\Delta)$. Thus, density of $I^*$ is $N_{[I^*]}=2.7\times
10^{14}~ cm^{-3}$. Recoil line width at $p=1~torr$, and radius of
the iodine atom $r_{[I]}=0.136~nm$ is estimated as
$\Gamma/(2\pi)={16 p\, r_{[I]}^2}/\sqrt{\pi m\, k_b T}= 0.7~ MHz$,
while the Doppler line width is
$2\,\sqrt{\ln2}\,\Delta_D/(2\pi)=0.27 ~GHz$.

Lasing condition can be satisfied, for instance, at laser cavity
length $L=50~cm$ and two identical mirrors of reflectivity
$R=\exp(-\sigma\, N_{[I^*]}\, L)=0.9$.

\begin{table}
\caption{P-T- odd polarizability, angle of polarization plane
rotation at an absorbtion length, and that for iodine laser,
calculated for the "splitting"(atom EDM) and "mixing" mechanisms
at $d_e=10^{-30}$, ${\mathcal E}=10^4~V/cm$ and $Q/\Delta
Q=10^{6}$.}
\label{tab1}       
\begin{tabular*}{\textwidth}{@{}l*{15}{@{\extracolsep{0pt plus12pt}}l}}
\br
mech. & $u$ &$v$& $\sigma$, $cm^2$& $\beta_{PT}$, $cm^3$&$\phi(L_{abs}), rad$ &$\phi_{las}, rad$ \\
\noalign{\smallskip}\hline\noalign{\smallskip}
mix. & 1 &0.0021& $6.7\times 10^{-18}$&$4.7\times 10^{-42}$ &$2.1\times 10^{-19}$ &$4.2\times 10^{-13}$\\
split. & 1&0.0021 & $6.7\times 10^{-18}$&$7.8\times 10^{-43}$&$3.4\times 10^{-20}$&$6.9\times 10^{-14}$\\
mix. & 2.5 &0.078&$ 2.2\times 10^{-19}$&$1.95\times 10^{-42}$&$2.7\times 10^{-18}$ &$5.4\times 10^{-12}$\\
split.& 2.5 &0.078&$2.2\times 10^{-19}$&$1.2\times 10^{-42}$&$1.6\times 10^{-18}$&$3.2\times 10^{-12}$\\
\br
\end{tabular*}
\end{table}

Increasing of the  detuning improves the signal (two last lines in
a Table 1)
 because it increases the ratio $f(u,v)/g(u,v)$.
 However, the cross-section decreases. To satisfy
 lasing condition one have to increase exited iodine atom density
 (one may increases of the mirror reflectivity  instead, but, as it will
 be discussed below, the more reflectivity the more difficult to reach small $\Delta
 Q/Q$).
For the Iodine atom density $N_{[I^*]}=10^{16}~cm^{-3}$ (pressure
is $p_{[I*]}=0.37~ torr$) estimate of  recoil line width give:
$\Gamma/(2\pi)=25~ MHz$ ($\Delta_D$ is the same as above). Lasing
condition is satisfied with the mirror reflectivity
$R=\exp(-\sigma\, N_{[I^*]}\, L)=0.9$. This demand high pressure
$p\sim 37 ~torr$ generators of singlet oxygen, discussed in
\cite{gener}.

\subsection{Two section system with the cesium vapor cell}
\begin{figure}
\vspace{0. cm} \hspace{2. cm}
  \includegraphics[width=5.cm]{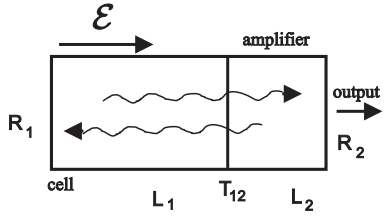}
 \caption{Simplified scheme of the two section
system.} \label{pin1}
\end{figure}

It does be desirable to realize an advantage of a strongly
forbidden transition.  For this aim one can use two section system
(Fig. \ref{pin1}), consisting of a cell with the cesium vapor in
the electric field and an amplifier. Let the length of the cell is
$L_1$ and the length of the amplifier is $L_2$. Circular
anisotropy takes place only in the cell, therefore the average P-
T- odd refractive index of system can be written down as $ \Delta
n _ {PT} \frac {L _ {1}} {L_1+L_2} $. The substance of the cell is
absorptive so the total absorption is written down similarly
(\ref{refmnim}) (see Appendix) as:
 \begin {equation}
\fl\frac {1} {2 Q} =-\frac {\ln (R_1T_{12}^2R_2e ^ {-2\,L_1/L _
{abs}})} {  4\,k (L_1+L_2)} = -\frac {\ln (R_1T_{12}^2R_2)} {4\,k
(L_1+L_2)} + \frac {L_1} {2\,k \, L _ {abs} (L_1+L_2)}, \label{19}
 \end {equation}
where $T_{12}$ is the transmittance of the wall between the
amplifier and the cell. The first item $ \frac {1} {2\,Q_0}=-\frac
{\ln (R_1T_{12}^2R_2)} {4\,k (L_1+L_2)}$ in (\ref{19}) describes
losses of the empty resonator.

\begin{figure}
\vspace{0 cm} \hspace{2 cm} \resizebox{0.48\textwidth}{!}{
  \includegraphics{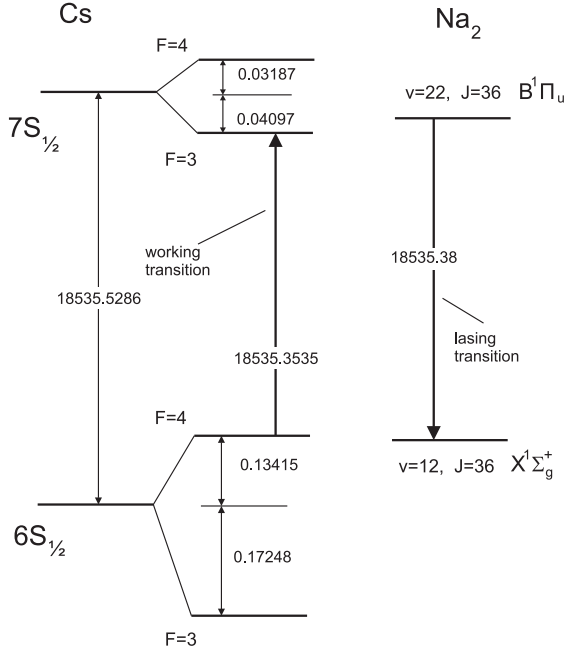}
} \vspace*{0. cm} \caption{Energies of Cesium working transition
and ${Na}_2$ lasing transition, $cm^{-1}$.}
\label{levels}       
\end{figure}

The angle of the polarization plane rotation in such two-section
system is
\begin {equation}
\phi _ {2 \, las} \sim {\Delta n _ {PT}}  \frac {L_1} {L_1+L_2}
\left ({\frac {1} {Q_0} + \frac {1} {k \, L _ {abs}} \frac {L_1}
{L_1+L_2}} \right) ^ {-2} \frac {1} {\Delta Q_0}, \label {2sek}
\end {equation}
where $\Delta Q_0$ describes the  linear anisotropy of losses of
the resonator. Corresponding lasing condition can be written as
\begin{equation}
2\kappa\, L_2- 2L_1/L_{abs}=-\ln(R_1 R_2 T_{12}^2), \label{las2}
\end{equation}
where $\kappa$ is pass gain constant of the amplifier and gives
$Q_0=\frac{k(L_1+L_2)}{\kappa\, L_2-L_1/L_{abs}}$.

 Now one may use the strongly forbidden magnetic transition
of the cesium atom  $6S _ {1/2} \rightarrow 7S _ {1/2} $. Let us
suggest sodium vapor containing molecules $Na_2$ as an amplifier.
Lasing line $539.4\pm 0.1~ nm$ of transitions
$B^1\,\Pi_u\rightarrow X^1\, \Sigma^+_g$ between vibrotational
levels of $Na_2$ molecule were found \cite{appl} among other
lasing lines under pumping $472.7~nm$ by Argon laser. The working
cavity contained sodium vapor at $820 ~K$, $p=11~torr$ and buffer
gas (Argon) with partial pressure $p_{[Ar]}=8~ torr$ \cite{appl}.
Dimer pressure was $p_{[Na_2]}\approx 0.05\, p$, corresponding to
the density of molecules $N_{[Na_2]}\approx 10^{16}~ cm^{-3}$.
Pass gain $\kappa=0.1~cm^{-1}$ was achieved. More accurately
energy of this lasing transition were measured by sodium vapor
spectroscopy \cite{spectr} to be $18535.38~cm^{-1}$ (the same
pumping was used).

The hyperfine structure of $6S_{1/2}-7S_{1/2}$ cesium transition
can be obtained  by compilation data of \cite{1,2,3} and is shown
in Fig. \ref{levels}. Nearest energy to that of lasing line has
the transition between hyperfine components $F=4$ and $F=3$.

The pressure of cesium vapor at the temperature $T=820 ~K $ is
$27~kPa$ (205 torr) and Cs atom density is $N=2.4\times 10 ^ {18}~
cm^{-3} $. Recoil line width estimated for cesium atom radius
$r_{[Cs]}=262~ nm$ is $\Gamma/(2\pi)= 0.33~ GHz$. Doppler line
width is $2\sqrt{\ln 2}\Delta_D/(2\pi)=1~GHz$. Thus parameters
$v=0.28$ and detuning
$u=\frac{\omega_{[Cs]}-\omega}{\Delta_D}=1.36$, where $\omega$ is
a frequency of ${Na}_2$ lasing transition and $\omega_{[Cs]}$ is
that for cesium working transition (Fig. \ref{levels}).
 Einstein coefficient
of the $Cs$ transition $6S_{1/2}\rightarrow 7 S_{1/2}$ in the
external electric field $10^4~ kV/cm$ is $A=0.034~s^{-1}$.
Absorbtion length is $L_{abs}=5~m$. Taking $L_1=30~cm$,
$L_2=10~cm$, according lasing condition (\ref{las2})  we find
$Q_0=5\times 10^6$, which can be realized with the two identical
mirrors of reflectivity $R=\exp(-\frac{Q_0}{k(L_1+L_2)})=0.34$,
where $T_{12}$ is set to unity for simplicity. According to the
 \cite{mats} the constant $B_{PT}^{mix}=5.8\times
10^{-34}~cm^3 s^{-1}$ at $d_e=10^{-30}~ e\, cm$ and external
electric field $\mathcal E=10~ kV/cm$. After removing relativistic
suppression multiplier $Z^2 \alpha^2=0.16$ we have the quantity
$B_{PT}^{mix}=3.6\times 10^{-33}~cm^3 s^{-1}$, which
 gives P-T- violating polarizability $\beta_ {PT}^{mix}=B^{mix}_{PT}\,{g(u,v)}/{\Delta_D}=7.1\times 10^{-43}~
 cm^3$.

Accordingly, the angle of polarization plane rotation in the
two-section system is $\phi_{2\,las}=7\times 10 ^ {-11} ~ rad $ at
$\Delta Q_0=10^{-6}Q_0$. However, the more is the number of
borders the more difficult is to achieve low residual anisotropy
of the system. How to avoid an additional border is discussed
below.

\subsection{Mixture of two gases}

In principle one may use mixture of gases\footnote{The idea was
suggested one of the referees of the article.} in a single laser
cavity. Concerning to the previous example this means that the
sodium laser have to work with the $205~torr$ cesium vapor instead
of $8~ torr$ Argon buffer gas. This can occur if the quenching of
the exited state of ${Na}_2$ molecule by cesium will not be strong
and should be checked experimentally (certainly one may diminish
cesium density to make laser functional). For the pass gain
$\kappa=0.1~cm^{-1}$, and $L=40~ cm $ the lasing condition is
satisfied with the mirrors reflectivity $R=\exp(-\kappa \,L)=0.02$
showing  that we have a reserve if the gain will be smaller due to
quenching.

 According to the eq. (\ref{sng}), (\ref{np}),
and previous subsection estimates for the cesium atom density and
linewidthes  we have $\phi_{las}=2.5\times 10^{-11}~ rad$.

Note that at present time sensitivity $10^{-8} rad/\sqrt{Hz}$ for
measurements of polarization rotation angle is achieved
\cite{cameron}. At accumulation time $10^6$ s (11.5 days) this
gives $10^{-11}$ rad.

Let us to do some remarks about residual anisotropy of resonator.
Suppose that the resonator consists of two identical mirrors and
anisotropy is created by the anisotropy of the mirror reflectivity
$\Delta R$. Thus we have $ \frac{\Delta Q}{Q}=-\frac{1}{\ln
R}\frac{\Delta R}{R}. $ This expression tells us that the more
reflectivity of the mirrors, the more ${\Delta Q}/{Q}$ at the same
${\Delta R}/{R}$. According to the expression $\phi_{las}=\Delta
n_{PT} Q^2/\Delta Q$ polarization plane rotation for the laser
with mixture of a gases can be rewritten in terms of ${\Delta
R}/{R}$ as $\phi_{las}=\Delta n_{PT}\,k\, L\frac{R}{\Delta R}, $
and for the case of two section system we have form (\ref{2sek}):
$ \phi_{2las}=\Delta n_{PT}\,k\, L_1\frac{R}{\Delta R}$
 when $k\, L_{abs}>> Q_0$. The above consideration shows, that
 because the working size of the resonator $L$ and $L_1$ are not be
 increased considerably due to presence of a strong electric field
 the only possibility to increase the effect is to lower anisotropy of
 loses $\Delta R$ of the mirrors.
Let us remind that the design of pumping should be done axially
symmetric.

\section {Conclusion}
We have considered P-T- odd rotation of the polarization plane of
the laser in the axial electric field to obtain new constraint to
the P-T- odd interactions.
 The main problems have to be solved are the
measurement of the tiny angle of polarization plane rotation and
producing the resonators with small linear anisotropy of losses.
Let us lay down some ways to this aim.

Concerning to the resonator, it can be made with movable mirrors
included to the self-consistent scheme of measurements to suppress
linear anisotropy of losses by fine tuning.

Then, to lower mean anisotropy of the mirror the technology can be
developed of producing the mirror itself as consisting from the
small sub-mirrors with chaotic orientation.

Then, it is desirable to  modulate the external electric field to
avoid possible systematic errors quadratic on the strength of the
field.

At last, as it has be done in the laser magnetometery
\cite{bret1}, the small external axial magnetic field can be
applied to compel the polarization plane to rotate. In such a way
the problem of the measurement of the small angles turns to the
problem of measuring of the frequency difference of polarization
plane rotation with and without electric field.

 To summarize, use of photon traps with an
amplifier for measurements of P-T- invariance violation with the
modern high technological level of the equipment (residual
anisotropy $\frac {\Delta Q} {Q} \sim 10 ^ {-6}$, external
electric field $10~kV/cm$ and ability to measure angles of
polarization rotation $~10^{-11}-10^{-12}~rad$ allows one to
achieve the sensitivity for electron dipole moment measurements at
the level of $10^{-30} ~e \times cm $ and, thus, to test the
predictions of some supersymmetric models. We hope that progress
in technology of resonators and precision polarization
measurements makes such experiments possible.

\section {Appendix}

One can obtain condition for stationary oscillation in a resonator
 supposing that the amplitude of the running wave after light travels
 back and forth in the resonator is equal to the initial amplitude.
 If, for example, the resonator is filled with a medium with a
 constant refractive index $n $ then the amplitude of a running wave
 departed from some point near the mirror (Fig. \ref{pin}) and returned to
 the original point will be equal to $ \vec E=e ^ {-2iknL} \vec E_0 $,
 where $k $ is the wave number of the wave in vacuum, $L $ is the length
 of the resonator. Since the initial and final amplitudes are equal to each
 other, $knL=m \,\pi $, where $m$ is an integer number. Difference of
 the refractive index from that given by the above condition by $
 \Delta n $ results in change of  the amplitude $ \Delta \vec E =-2i\Delta
 n \, k \, L \,\vec E $ after the full passage.  Dividing the last
 equation by the propagation time $T\approx 2L/c$ we find:

\begin {equation}
\frac {d\vec E} {d t} =-i\Delta n \, \omega \vec E. \label {sh}
\end {equation}
This equation may be derived also in a less heuristic way
\cite{lamb,gal,harr,gal1,voit}.  Losses corresponding to the
reflectance of the resonator mirrors can be ``smeared out'' over
the volume of the resonator through the addition of quantity:
\begin{equation}
 \frac {i} {2\,Q} = -\frac {i} {2} \frac {\ln (R_1R_2)} {2 k \,L}
\label{refmnim}
\end{equation}
to $ \Delta n $, where $R_1, R_2$ are the reflectance of mirrors.
It is easy to see that after the full pass the amplitude of the
wave is multiplied by $\sqrt{R_1R_2} $. If there are some areas
with small variations of the refractive index then we should use
the average index of refraction $ \Delta n = \frac {L_1\Delta
n_1+L_2\Delta n_2 +\dots} {L_1 +L_2 +\dots}$.

If the wave propagates along $z$ axis then the amplitude of the
electromagnetic wave contains two components $ \vec E = \left (
\begin {array} {c}
E_x \\
E_y
\end {array} \right) $.
It is convenient to use the circular basis $E _ + =-\frac {1}
{\sqrt {2}} (E_x+iE_y) $, $E_-=\frac {1} {\sqrt {2}} (E_x-iE_y) $.
In the general case of an anisotropic medium and resonator, $
\Delta n $ is a complex $2\times2 $ matrix which can be written in
the form $ \hat \Delta n = ({\mathcal N}_0 +\vec \sigma \vec
{\mathcal N}) $, where $\mathcal N_0=r_0+ih_0 $ is a complex
number, $ \vec {\mathcal N} =\vec r+i\vec h $ is a complex vector,
$ \vec \sigma \equiv \{\sigma_x, \sigma_y, \sigma_z \} $ are the
Pauli matrices.

 Let us define the density matrix $ \rho _
{ij} (t) =E_i (t) E ^ * _ j (t) $ and parameterize it with the
help of $ \xi_0 $ and $ \vec \xi $ as $ \rho = (\xi_0 +\vec \sigma
\vec \xi)/2 $. Taking derivatives of the density matrix and
replacing derivatives of $\frac{d E_j}{d t}$ using (\ref {sh}) we
find $i\frac {1} {\omega} \frac {d\rho} {dt} = \Delta \hat
n\rho-\rho \Delta \hat n ^ + $, which results in
\begin {eqnarray}
\frac {1} {2\omega} \frac {d\vec \xi} {d t} =h_0\vec
\xi +\xi_0\vec h +\vec r\times\vec \xi,\nonumber \\
\frac {1} {2\omega} \frac {d\xi_0} {d t} =h_0\xi_0 + (\vec
h\cdot\vec \xi). \label {eqxi}
\end {eqnarray}

From the definition of a density matrix  we see, that
\begin {eqnarray}
\xi_x = <\sigma_x > =E ^ * _ + E_-+ E_-^*E _ +, ~~\nonumber \\
\xi_y = <\sigma_y > =-i (E _ + ^*E_ - - E_-^*E _ +),\nonumber \\
\xi_z = <\sigma_z > =E ^ * _ + E _ +-E ^ *_-E_-,\nonumber \\
\xi_0 = <I> =E ^ * _ + E _ ++ E ^ *_-E_-,
\end {eqnarray}
where $I$ denotes unit matrix. Instead of $ \vec \xi $ and $ \xi_0
$ let's define the unit vector $\vec O =\frac {\vec \xi} {\xi_0}
$. Equations (\ref {eqxi}) then take the form
\begin {equation}
\frac {1} {2\omega} \frac {d\vec O} {dt} = \vec r\times\vec O
+\vec h-\vec O(\vec h\cdot\vec O). \label {eqO}
\end {equation}
For entirely polarized light $|\vec O|=1$ the Eq.(\ref{eqO})
reduces to
\begin {equation}
\frac {1} {2\omega} \frac {d\vec O} {dt} = \vec r\times\vec O
+\vec O\times (\vec h\times\vec O).
\end {equation}

 Orientation of the polarization ellipse is described by the
angle $\phi $ between the major axis of  polarization ellipse and
$x$ \cite{landay}:
\[ \mbox{tg} \, 2\phi =\frac {E_yE_x^* + E_xE^*_y} {E_xE_x
^*-E_yE_y ^ *} =-i\frac {E _ + E_-^*-E_-E ^ * _ +} {E _ + E ^ *_-+
E _-E _ + ^ *} =-\frac {O_y} {O_x}.
\]
Thus the angle of polarization rotation is equal to one half of
the angle of the vector $ \vec O_\perp $ rotation in a plane $xy
$, taken with the opposite sign. For example, if $ \vec O_\perp $
rotates on the angle $\left(-\frac {\pi}{2}\right)$, then the
ellipse of polarization rotates on the angle $ \frac {\pi} {4} $.
Rotation of $ \vec O_\perp $ by the angle $\left(-2\pi\right) $
means rotation of the polarization ellipse on the angle $ \pi $
and, after this rotation the ellipse coincides with itself.

The z-component of the vector $ \vec O $ is equal to the
ellipticity of laser radiations.

\begin{figure}
\vspace{0 cm} \hspace{2 cm} \resizebox{0.48\textwidth}{!}{
  \includegraphics{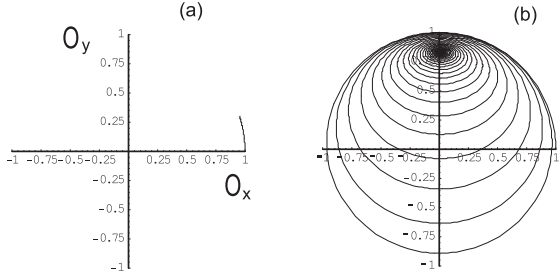}
} \vspace*{0. cm} \caption{Evolution of polarization toward
stationary solution in the case $\vec r=\{0.01, 0, 0.3\}, \vec
h=\{1, 0, 0\}$,(a) and when $\vec r=\{0.01, 0, 1.2\}, \vec h=\{1,
0, 0\}$ (b)} \label{evol}
\end{figure}

In the general case  $ (\vec h\cdot\vec r) \neq 0 $ polarization
always tends  to the stationary solution. The typical picture of
this is shown in Fig. \ref{evol}. Under constant $ \vec h $ and $
\vec r $ the stationary solution of the equation (\ref {eqO}) is
written down as
\begin {eqnarray}
\vec O_0 =\alpha \,\vec r +\beta \,\vec h +\gamma \,\vec
r\times\vec h,
\nonumber \\
\nonumber \\
\gamma =\frac {h^2+r^2-\sqrt {(h^2+r^2) ^2-4\mid\vec h\times\vec r
\mid^2}} {2\mid\vec h\times\vec r
\mid^2}, \nonumber \\
\alpha =\pm\sqrt {\gamma (1-\gamma h^2)}, ~~\beta =\pm\frac
{\gamma ^ {3/2} (\vec r\cdot\vec h)} {\sqrt {1-\gamma h^2}}.
\label {sol}
\end {eqnarray}
Signs  $ \pm $ correspond  to two various solutions. One of them
is stable.  In the particular case  $ (\vec h\cdot\vec r) =0 $ the
solution (\ref {sol}) reduces to
\begin {equation}
\fl\vec O_0 =\pm \frac {\sqrt {h^2-r^2}} {h^2} \vec h +\frac {\vec
r\times \vec h} {h^2}, ~~\mid\vec h\mid > \mid \vec r\mid;~~
 \vec O_0
=\pm \frac {\sqrt {r^2-h^2}} {r^2} \vec r +\frac {\vec r\times
\vec h} {r^2},~~\mid\vec r\mid > \mid \vec h\mid. \label{formap}
\end {equation}

Nonlinear properties of the laser medium  results in dependence of
the  factor $h_0 $ on $ \mid E \mid^2 =\xi_0$. In this case in
addition to the equation (\ref{eqO}) we should consider equation
\begin {equation}
\frac {1} {2\omega} \frac {d\ln\xi_0} {dt} =h_0 + (\vec h\cdot\vec
O).\label{eqq}
\end {equation}
However, nonlinearity of this type has no influence on the
polarization evolution.

One more manifestation of nonlinearity is so-called
"self-rotation" \cite{gal1} resulting in possibility of
dependence:
\begin {equation}
\frac{i}{\omega}\frac {d\vec E} {dt} \sim a _ {sf} (\vec
E\cdot\vec E)\vec E ^*.
\end {equation}
For $E_-$ and $E _ + $ we have:
\begin {eqnarray*}
\fl\frac{i}{\omega}\frac {dE_-} {dt} \sim\frac {i} {\omega\sqrt
{2}} \left ( \frac {dE_x} {dt}-i\frac {E_y} {dt} \right) \sim\frac
{a_{sf}} {\sqrt {2}} E^2 (E_x ^*-iE_y ^ *) \sim
-a_{sf}E^2E _ + ^*\sim 2 a_{sf}\mid E _ +\mid^2 E_-,\nonumber \\
\fl\frac{i}{\omega}\frac {dE _ +} {dt} \sim-\frac {i} {\omega\sqrt
{2}} \left ( \frac {dE_x} {dt} +i\frac {E_y} {dt} \right)
\sim-\frac {a_{sf}} {\sqrt {2}} E^2 (E_x ^ * + iE_y ^ *)\sim
-a_{sf}E^2E_-^*\sim 2 a_{sf}\mid E_-\mid^2 E _ +,
\end {eqnarray*}
(we use here the expression $E^2 =-2E_-E _ + $). Taking into
account that $ \mid E _ +\mid^2 =\frac {\xi_0} {2} (1+O_z) $ and $
\mid E_-\mid^2 =\frac {\xi_0} {2} (1-O_z) $ we can see the terms
$r_z\sim-\mbox {Re}\, a _ {sf} \xi_0 O_z $, $h_z\sim-\mbox {Im}\,
a _ {sf} \xi_0 O_z $ appearing in the z-components of $\vec r $
and $\vec h$. This results in rotation of the plane of
polarization even in the absence of circular anisotropy. However,
rotation does not depends on external electric field and can be
distinguished from the P-T- noninvariant effects by modulation of
the external electric field, moreover, the ellipticity $O_z $ of
the laser light is extremely low.

\end{document}